\documentclass[pre,a4paper,superscriptaddress,twocolumn,showpacs,amsmath,amssymb,floatfix]{revtex4}

\usepackage{graphicx}
\usepackage{amssymb}
\usepackage{float}

\usepackage{color}

\begin{document}
	
\title{Deconfinement of classical Yang-Mills color fields
       in a disorder potential}
	
	\author{Leonardo Ermann}
	\affiliation{Departamento de F\'{\i}sica Te\'orica, GIyA,
         Comisi\'on Nacional de Energ\'{\i}a At\'omica.
           Av.~del Libertador 8250, 1429 Buenos Aires, Argentina}
         \affiliation{Consejo Nacional de Investigaciones
             Cient\'ificas y T\'ecnicas (CONICET), Buenos Aires, Argentina}
	\author{Dima L. Shepelyansky}
	%\homepage[]{http://www.quantware.ups-tlse.fr}
	\affiliation{\mbox{Laboratoire de Physique Th\'eorique, 
			Universit\'e de Toulouse, CNRS, UPS, 31062 Toulouse, France}}
	
	%\date{today}
	\date{March 30, 2021}
	
	\begin{abstract}
	We study numerically and analytically the behavior of classical Yang-Mills color fields
        in a random one-dimensional potential described by
        the Anderson model with disorder. Above a certain threshold 
         the nonlinear interactions of Yang-Mills fields lead to chaos and deconfinement 
        of color wavepackets  with their subdiffusive spreading in space.
        The algebraic exponent of the second moment growth in time is found to be in a range 
        of 0.3 to 0.4. Below the threshold  color wavepackets remain confined
        even if a very slow spreading at very long times is not
        excluded due to subtle nonlinear effects and the Arnold diffusion
        for the case when initially color packets are located in a close vicinity.
         In a case of large initial separation of color wavepackets 
         they remain well confined and localized in space.
         We also present comparison with the behavior of the one-component field model of 
         discrete Anderson nonlinear Schr\"odinger equation with disorder.
	\end{abstract}

	%\pacs{05.45.Mt, 67.85.Hj,  47.27.-i, 72.15.Rn}
	%05.45.-a Nonlinear dynamics and chaos
	%67.85.Hj 	Bose-Einstein condensates in optical potentials
	%47.27.-i 	Turbulent flows
	%72.15.Rn 	Localization effects (Anderson or weak localization) 
	%
	%47.35.-i 	Hydrodynamic waves
	%47.35.Bb 	Gravity waves 
	%89.75.-k 	Complex systems 

	\maketitle
	
	%{\it Introduction.} 
	\section{Introduction} 
	\label{sec1}

The Yang-Mills (YM) gauge fields were introduced \cite{ym}
for an isotropic-invariant description of strong interactions.
The investigation of properties of these fields
still remains an interesting and important problem.
The studies of classical YM fields are also important for
applications in several problems of quantization \cite{polyakov1,polyakov2}.
The classical dynamics of these fields is essentially nonlinear and nontrivial.
Its analysis is rather important for semiclassical description of 
strong YM vacuum fluctuations \cite{polyakov3,polyakov4,vainstein,shuryak2002}.
Thus the investigation of nonlinear dynamics and time evolution 
of classical YM fields represents a relevant topic.
 
The important class of classical YM models was introduced
in \cite{matinyan1} where the YM fields are homogeneous
in space so that the time evolution is described only by nonlinear dynamics of 
interacting colors. In general this Hamiltonian dynamics of color YM fields
was shown to be chaotic \cite{chis1,matinyan2,chis2}  
even if certain integrable solutions also exist.
Thus the  YM dynamics belongs to a generic class
of chaotic Hamiltonian systems with divided phase space 
with small integrability islands embedded in a chaotic sea \cite{chirikov1979,lichtenberg}.
Even if important mathematical results have been obtained for 
 chaotic dynamics (see e.g. \cite{arnold,sinai})
the properties of chaos  with such a divided phase space,
composed of integrable islands surrounded by a chaotic component, 
still remain very difficult for mathematical analysis.
The existence of chaos of classical homogeneous YM fields has been reported 
already some time ago  \cite{matinyan1,matinyan2,chis1,chis2} but still
these YM fields and related models attract attention of 
researchers (see e.g. \cite{ymclass1,ymclass2,ymclass3}).

The above dynamics of YM color fields can be reduced to a rather simple Hamiltonian
which for $N_C=2,3$ colors reads:
\begin{equation}
%\begin{array}{cll}
H= \sum^{N_C}_{i=1} ({p_\mu}^2 + m{x_\mu}^2)/2 + \beta \sum_{\mu' \neq \mu} {x_{\mu'}}^2 {x_\mu}^2/2  \; ,  
%\end{array}
\label{eq_3colorham}
\end{equation}
where $(p_\mu, x_\nu)$  are effective conjugated momentum and coordinate, 
color index is $\mu=1, ... N_C$,
 $m$ is mass, which is zero or finite in presence of Higgs mechanism, 
and $\beta$ determines the strength of nonlinear interactions 
of colors \cite{matinyan1,chis1,matinyan2,chis2}.
An interesting feature of the finite mass case (e.g. $m=1$ in dimensionless units used here) is that
the measure of chaos remains finite and large (about $50\%$) even in the limit 
of very weak nonlinearity $\beta \rightarrow 0$
since the Kolmogorov-Arnold-Moser (KAM) theorem \cite{lichtenberg} is not valid 
when all color masses (or oscillator frequencies) are the same \cite{chis2}. 

Till present the classical dynamics of YM colors
was analyzed for fields homogeneous in space. 
Here we consider the case of space nonhomogeneous  fields.
Namely, we  study a spreading of such YM fields
in space in presence of disorder potential
which corresponds to another generic limiting case of space properties.
Such a disorder corresponds to  random properties of 
vacuum in Quantum Chromodynamics (QCD) 
discussed in the literature (see e.g. \cite{shuryak1980,olesen,kirzhnits,shuryak1993}).
It is well known that in quantum mechanics a disorder potential 
may lead to a localization of probability spreading 
due to quantum interference effects. This phenomenon
is known as the Anderson localization \cite{anderson}
and plays an important role for electron transport in
solid-state systems with disorder \cite{imry,montambaux,mirlin}.
The eigenstates of such a system are exponentially localized 
in 1 and 2 dimensions (1D and 2D) while in 3 dimensions (3D)
a delocalization transition takes place
at a disorder below certain threshold (see e.g. review \cite{mirlin}). 

The effects on nonlinearity on Anderson localization in 1D lattice
were investigated in \cite{dls1993} where it was shown that
the localization is preserved at weak nonlinearity
while above a certain threshold a subdiffusive 
spreading over the whole lattice takes place.
The detailed numerical studies of this phenomenon
in Disordered Anderson Nonlinear Schr\"odinger Equation (DANSE)
have been reported in \cite{molina,dls2008,flach2009} and results of different groups 
were reviewed in \cite{mulansky,flach}. The subdiffusive spreading
has been studied for various  nonlinear models in 1D and 2D
(see e.g. \cite{garcia,flachkg,ermann,flach2019,skokos}) with a  spreading
continuing up to enormously long dimensional times $t \sim 2 \times 10^{12}$ 
reported for a 1D model in \cite{flach2019}. The interest to the effects of
nonlinearity on Anderson localization is also supported by
related experimental studies of wave propagation 
in a disordered nonlinear media \cite{segev,lahini} and spreading of
Bose-Einstein cold atom condensates in optical disorder lattices \cite{inguscio1,inguscio2}
described by the Gross-Pitaevskii equation.

All above investigations of packet spreading in a disorder potential with nonlinearity
have been done for one-component nonlinear field of DANSE
with nonlinear self-interaction (see e.g. \cite{dls2008,mulansky,flach}).
The case of YM color dynamics is different since nonlinearity
appears only due to interactions of color components.
In fact possible implications of randomness, dynamical chaos, 
Anderson localization and confinement has been discussed in \cite{olesen,kirzhnits}.
The deconfinement transition in QCD at finite temperature is also
under active investigation (see e.g. \cite{deconf1,deconf2,deconf3} and Refs. therein). 
Here,  we find that under certain conditions the nonlinear interaction of YM colors
leads to deconfinement of YM fields and their unlimited subdiffusive spreading
in space. In the case of weak nonlinearity or spacial separation of YM color 
components the Anderson localization is preserved and fields remain
localized in space. We hope that the obtained results
may be of interest for the deconfinement phenomenon of quantum YM fields 
which attracts a significant interest.

The paper is organized as follows: in Section~\ref{sec2}, we give the system description,
the numerical and analytical results are  given in Section~\ref{sec3}, 
discussion of results is given in Section~\ref{sec4}.
	
%{\it Model description.}
\section{Model description} 
\label{sec2}

\subsection{DANSE}

Me start with a brief description of DANSE model
studied in \cite{dls2008,mulansky,flach,garcia}.
The wavefunction evolution of DANSE is described by the equation:
\begin{equation}
i \hbar{{\partial {\psi}_{n}} \over {\partial {t}}}
=E_{n}{\psi}_{n}
+{\beta}{\mid{\psi_{n}}\mid}^2 \psi_{n}
 +V ({\psi_{n+1}}+ {\psi_{n-1})} \;.
\label{eq_danse}
\end{equation}
Here 
$\beta$ determines nonlinearity strength,
$V$ gives near-neighboring hopping matrix element, on-site
disorder energies are randomly distributed in the range
$-W/2 < E_n < W/2$, and the total probability is 
conserved and normalized to unity
$\sum_n {\mid{\psi_{n}}\mid}^2 =1$. For $\beta=0$ 
all eigenstates are exponentially localized
with $|\psi| \propto \exp(-|n-n_0|/\ell)$
and localization length is $\ell \approx 96 (V/W)^2 $
at the  energy band center  and weak disorder \cite{kramer1993}.
Here $n_0$ marks  a center of wavefunction. 
We consider a case of relatively weak disorder with $\ell > 1$.
For convenience we set $\hbar=V=1$ so that 
the energy coincides with the frequency. 

Above a certain threshold $\beta > \beta_c$ the nonlinearity
leads to a destruction of localization
with a subdiffusive spreading of wavepacket width 
$\Delta n = n - n_0$:
\begin{equation}
\sigma = <(\Delta n)^2 > \propto t^\alpha  \; ,
\label{eq_spread}
\end{equation}
where  brackets mark averaging over wavefunction at time $t$
and $\alpha$ is the subdiffusion exponent.
The numerical simulations give its value being
in a range $0.3 \leq  \alpha \leq 0.4$.
Certain analytical arguments were given for values $\alpha=0.4$ \cite{dls1993,dls2008,garcia}
and $\alpha = 1/3$ \cite{flach}. An introduction
of randomness in eigenstate phases of linear problem
is supposed to produce a spreading with $\alpha=0.5$ \cite{basko,flach}.
Indeed, an increase of dephasing leads to a growth of $\alpha$
approaching the value $\alpha=0.5$ \cite{ermann}.

It is difficult to give an exact estimate of the threshold value $\beta_c$. 
The numerical results show that at $\beta =0.1; 0.03$ the wavepacket square width $\sigma$
remains bounded without significant 
increase up to times $t=10^8$ \cite{dls2008}. 
However, it is possible that some type of Arnold diffusion 
along tiny chaotic layers \cite{chirikov1979,lichtenberg,chivech} 
may lead to a very slow spreading of a very small wavepacket fraction. 
It should be pointed that the Anderson localization
is characterized by a pure-point dense spectrum and its
perturbation by nonlinearity represents a very difficult
problem for mathematical analysis.
A reader can find some mathematical results
for this problem reported in \cite{fishman,wang}.

A surprising feature of unlimited spreading at $\beta > \beta_c$
is that with growth of $\Delta n$ the relative local
contribution of nonlinear term in (\ref{eq_danse})
decreases as $\beta |\psi|^2 \sim \beta/\Delta n \propto \beta /t^{\alpha/2}$
and on a first glance it seems that nonlinearity
becomes weaker and weaker with time. In \cite{dls1993}
is was argued that even being small this term gives a local nonlinear frequency
spreading $\delta \omega \sim  \beta/\Delta n$ which 
at $\beta > \beta_c \sim 1$
remains larger than the typical spacing $\Delta \omega \sim 1/\Delta n$ 
between frequencies of linear eigenmodes populated due to 
subdiffusive spreading of wavepacket at time $t$.
As soon as $\delta \omega > \Delta \omega$   
the spectrum of motion  remains continuous and 
thus the spreading can continue unlimitedly in time.
However, a better understanding of origins of such
unlimited spreading is still highly desirable.

\subsection{YM color models}

In a similarity with dynamics of homogeneous YM fields described by  Hamiltonian 
(\ref{eq_3colorham}) and DANSE (\ref{eq_danse}) we model the dynamics of YM color fields
in a disorder potential by the nonlinear Schr\"odinger equation:
\begin{equation}
i {\partial {\psi^{\mu}_{n}} \over {\partial {t}}}
=E_{n} \psi^{\mu}_{n}
+{\beta}({\sum_{\mu' \neq \mu} \mid{\psi^{\mu'}_{n}}\mid}^2) \psi^{\mu}_{n}
 + ({\psi^{\mu}_{n+1}}+ {\psi^{\mu}_{n-1})} \;.
\label{eq_ym}
\end{equation}
Here $\mu =1,...,N_C$ is color index changing from $1$ to $2$ for two YM colors
$N_C=2$ or from $1$ to $3$ for three colors $N_C=3$.
We denote these two cases as $YMCA2$ and $YMCA3$ respectively (with A for agent).
At zero nonlinearity $\beta=0$ each color
evolution is described by 1D Anderson model with the same disorder $E_n$ for all colors
and being the same as in (\ref{eq_danse}). In absence of hopping to nearby sites
and all energies $E_n$ being equal we have the dynamics of color fields
described by equations  similar to those for 
the homogeneous YM fields from Hamiltonian (\ref{eq_3colorham}). 
Thus we consider the equations (\ref{eq_ym}) as a realistic
model of evolution of classical YM color fields in a disorder potential.

\begin{figure}[t]
\begin{center}
\includegraphics[width=0.46\textwidth]{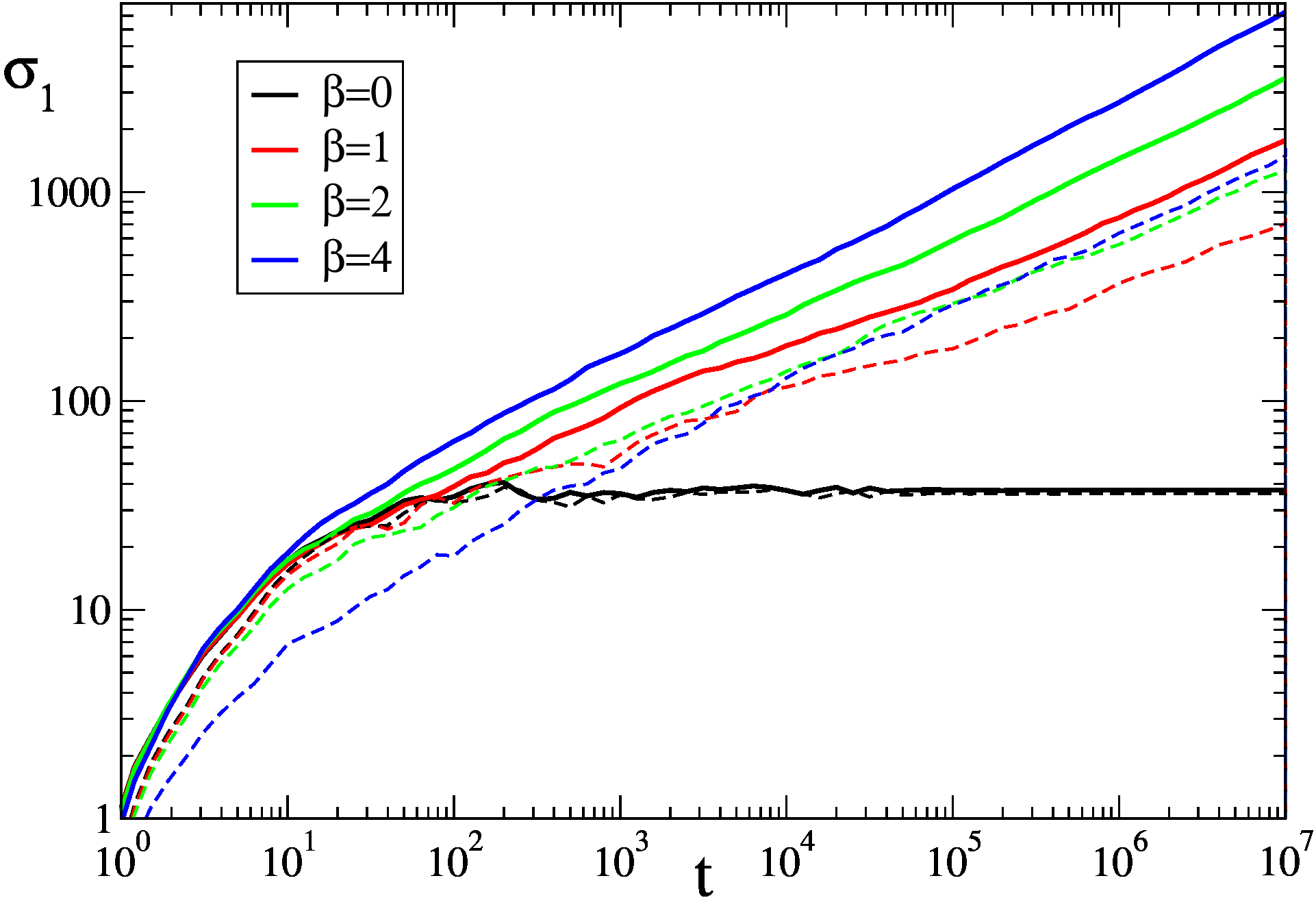}
\end{center}
\vglue -0.3cm
\caption{\label{fig1}
Time evolution of the second moment $\sigma_1$ of probability distribution, defined in the text, 
for YMCA3 model with 3 colors (\ref{eq_ym}) (solid curves) and
DANSE model (\ref{eq_danse}) (dashed curves) at 
$\beta=0$ (black curves), 
$\beta=1$ (red curves), $\beta=2$ (green curves) and $\beta=4$ (blue curves)
at $W=4$. 
At initial time the 3 color packets are located at 3 different sites $n=-1,0,1$ for YMCA3;
for DANSE initial probability is at $n=0$.
The average is done over 20 random realizations of disorder and 
over logarithmic equidistant intervals of time. 
}
\end{figure}

As for DANSE, the evolution of YM fields (\ref{eq_ym}) has the energy 
conservation, also the probability is conserved for each component 
normalized to unity $\sum_{n} \mid{\psi^{\mu}_{n}}\mid^2 =1$.
The numerical simulations of DANSE and Klein-Gordon nonlinear (KGN) model with disorder
\cite{flach,flachkg,ermannnjp} show that the exponent $\alpha$ 
is approximately the same in these two models even if
only energy is conserved in the KGN case. Thus we also expect
that the probability conservation for each color component will not affect
the spreading exponent $\alpha$.   Indeed, the number of degrees of freedom in (\ref{eq_ym})
is given by number of lattice sites multiplied by $N_C$ being much larger
than the number of integrals $N_C + 1$ of energy and component probabilities. 

\begin{figure}[t]
\begin{center}
\includegraphics[width=0.46\textwidth]{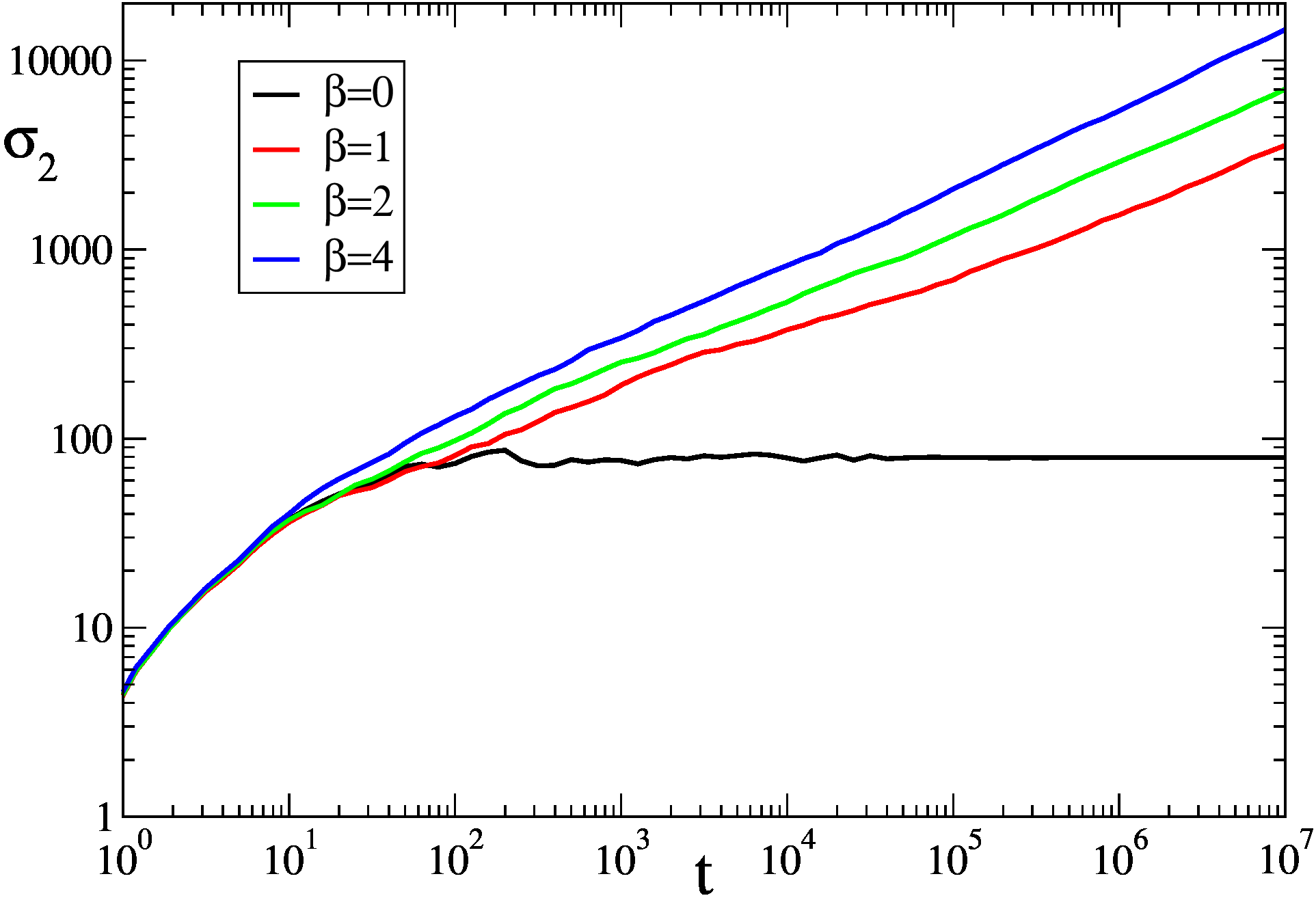}
\end{center}
\vglue -0.3cm
\caption{\label{fig2}
Same as in Fig.~\ref{fig1} but for the second moment $\sigma_2$ 
of probability distribution for YMCA3 model, defined in the text.
}
\end{figure}	

From the structure of YM equations (\ref{eq_ym}) we can make certain direct 
observations. At first, it is possible to consider the symmetric case when initially
all color components $\psi^{\mu}_{n}$ are the same. Then their
evolution is described by the DANSE equation (\ref{eq_danse})
with some rescaling of $\beta$ for $N_C=3$. However, since the field evolution
is chaotic this solution is unstable and small corrections 
to this symmetric state grow exponentially with time
so that this symmetry is completely destroyed very rapidly.
Still such a symmetric case allows to expect
that the spreading exponent $\alpha$
will have a value similar to those found for DANSE.
As for DANSE we expect that YM fields remain confined or localized
below a certain chaos threshold with $\beta \ll \beta_c \sim 1$.
In spite of this possible similarity between DANSE and YMCA models
there are two important differences between them.
Thus if initially color wavepackets are located far from each other
with a typical distance between them $R_C$
being significantly larger than the localization length
$\ell$ of the linear case ($R_c \gg \ell$) 
then an effective interaction between colors
becomes exponentially small $\beta_{eff} \propto \beta \exp(-2R_C/\ell)$
so that we have $\beta_{eff} \ll \beta_c \sim 1$.
Thus we expect that such initial states will remain 
exponentially localized or confined for all times.
Another new element of YMCA, compared to DANSE case, 
is that the eigenenergies $\varepsilon_m$ of linear problem eigenmodes  
at $\beta=0$ are the same for all colors.
Thus for one site and 3 colors we have a dynamics being very similar to 
those of Higgs case with finite mass (\ref{eq_3colorham})
studied in detail in \cite{chis2}. Due to this degeneracy
the KAM theorem cannot be applied to this system
and the measure of chaos remains about 50\% even 
in the limit of nonlinearity going to zero \cite{chis2}.
However, the initial wavepackets of colors should 
populate the same linear eigenmodes 
(this requires $R_C < \ell$).
Such situation also generally appears  in other type on nonlinear systems 
with many degrees of freedom \cite{mulansky2}.
Since in YMCA at $N_C=3$ (\ref{eq_ym}) 
there many eigenenergies $\varepsilon_m$ (linear frequencies)
which are the same we expect that there are many initial configurations
when colors are initially located on a distance $R_c \sim \ell$ and their
dynamics remains chaotic even for very small nonlinearity $\beta \rightarrow 0$.
However, a question about spreading of such chaos over lattice sites
remains open. 

\section{Numerical results for time evolution of YM colors}
\label{sec3}

Following the approach used in \cite{dls2008}, the
numerical integration of Eqs.~(\ref{eq_danse}),~(\ref{eq_ym})
is done by the Trotter decomposition with a time step $\Delta t =0.05$
and the total number of sites $N=1001$ for each color
with the fast Fourier transform from coordinate to momentum 
representation and back. This integration scheme is symplectic and conserves 
probability exactly. Its efficiency has been confirmed
by various numerical simulations (see e.g. \cite{dls2008,mulansky,flach,garcia}).
We checked that the variation of system size $N$ and integration time step $\Delta t$
does not affect the results. We present here the results mainly for 
a typical disorder strength $W=4$ and nonlinearity values $\beta=0,1,2,4$.
The spreading of color probabilities is characterized by 
the squared wavepacket width at different times defined as:
$\sigma_1 =   \langle n_1^2\rangle-\langle n_1\rangle^2$ for DANSE,
$\sigma_1 =\sum^{N_C}_{\mu=1} \left(\langle n_\mu^2\rangle-\langle n_\mu\rangle^2\rangle\right)/N_C$
for YMCA2, YMCA3 and relative square moments
$\sigma_2 = \left\langle(n_1-n_2)^2 \right\rangle$ for YMCA2 and
 $\sigma_2 = \left[\left\langle(n_1-n_2)^2 \right\rangle+
\left\langle(n_1-n_3)^2 \right\rangle+\left\langle(n_2-n_3)^2 \right\rangle\right]/3$ 
for YMCA3. Here brackets mark the average over wavefunction.
The results are also averaged over 20 disorder realisations. 

\subsection{Deconfinement and subdiffusive spreading of YM colors}

\begin{figure}[t]
\begin{center}
\includegraphics[width=0.46\textwidth]{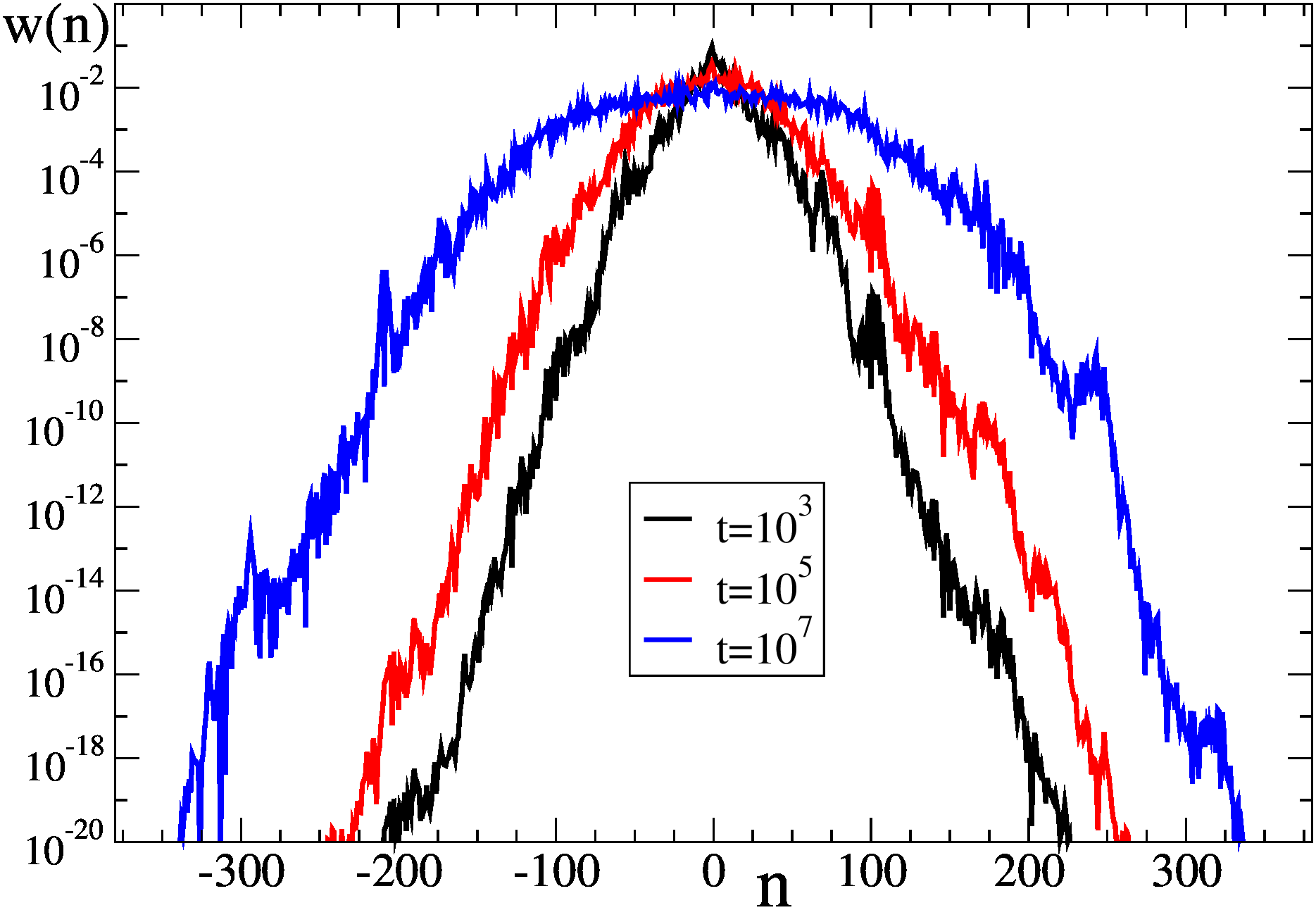}
\end{center}
\vglue -0.3cm
\caption{\label{fig3}
Probability distributions $w(n) = \sum_{\mu=1}^{N_C} \mid{\psi^{\mu}_{n}}\mid^2/N_C$ 
for YMCA3 model ($N_C=3$, $\beta=2$, $W=4$) are shown at times $t=10^3$ (black curve), $t=10^5$ 
(red curve) and $t=10^7$ (blue curve); 
probabilities are averaged over 20 disorder realizations. 
}
\end{figure}

\begin{figure}[t]
\begin{center}
\includegraphics[width=0.46\textwidth]{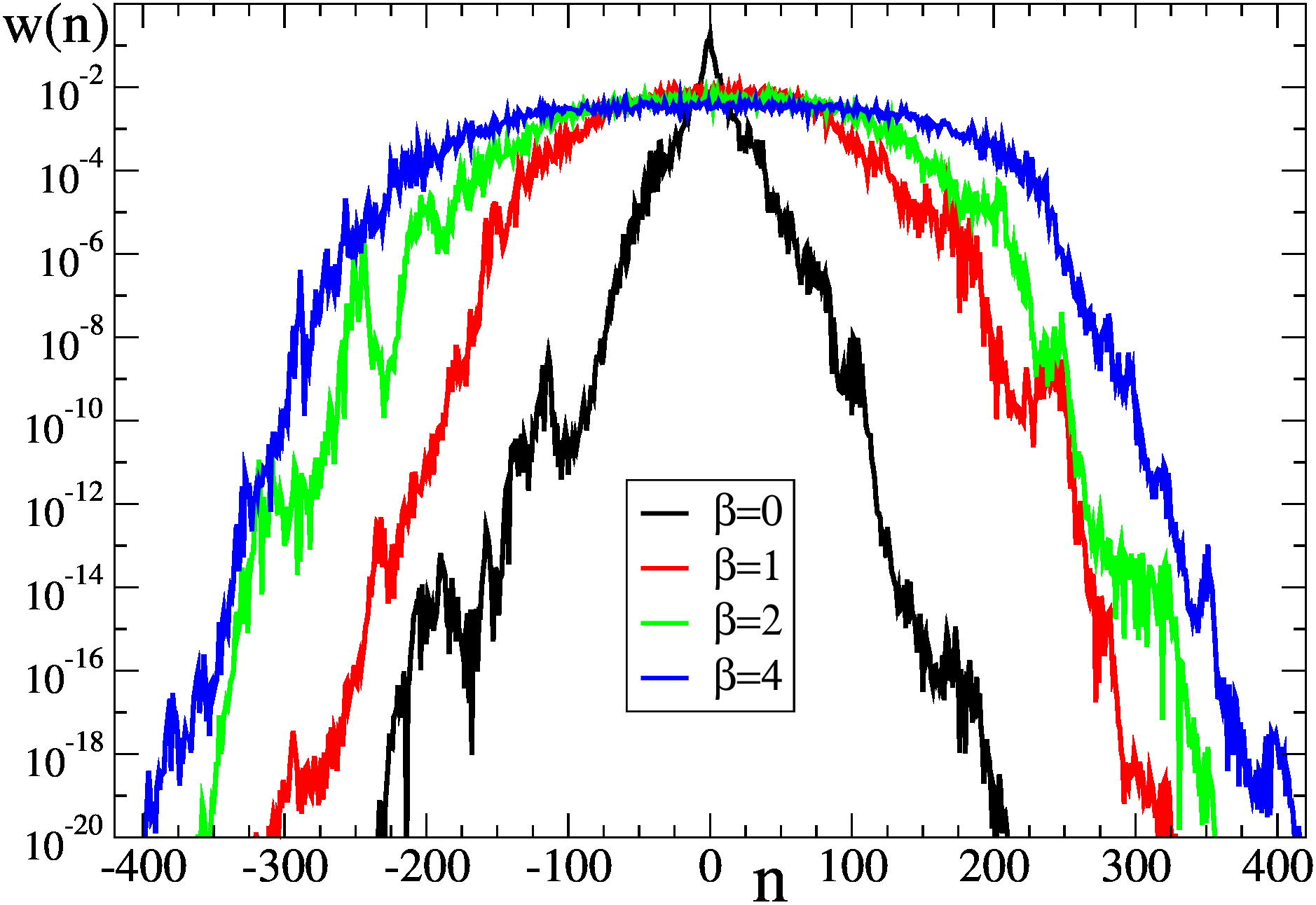}
\end{center}
\vglue -0.3cm
\caption{\label{fig4}
Same as in Fig.~\ref{fig3} but all distributions $w(n)$ of YMCA3
are shown at time $t=10^7$ for $\beta=0$ (black curve), 
$\beta=1$ (red curve), $\beta=2$ (green curve) and $\beta=4$ (blue curve);
 probabilities are averaged over 20 disorder realizations.
}
\end{figure}

The time dependence of second moments $\sigma_1$ for DANSE and YMCA3 models
is shown in Fig.~\ref{fig1} for different values of $\beta$ and disorder $W=4$.
At such a disorder and $\beta=0$ the wavepacket spreads on approximately
$\Delta n \approx 7 $ sites in agreement with the theoretical
value of the localization length $\ell = 96/W^2 =6$.  In presence of nonlinear
interactions there is a subdiffusive spreading of wavepacket
which is somewhat stronger for YMCA3 compared to DANSE case.
The time evolution of the second moment $\sigma_2$ for YMCA3
case is shown in Fig.~\ref{fig2} for the same values of $\beta$ as in Fig.~\ref{fig1}.
The growth of both moments $\sigma_1$ and $\sigma_2$ is very similar.
This means that the color packets spread in such a way that they
remain close to each other so that their effective interactions 
allow to make correlated joint transitions over
localized eigenstates of the Anderson model at $\beta=0$.
It is clear that interactions of colors leads to deconfinement of YM fields
with the unlimited subdiffusive spreading over the whole lattice.
The growth of moments $\sigma_1, \sigma_2$ for YMCA2 case is very similar
to those of YMCA3 and we do not show it here (but the obtained
spearing exponents $\alpha$ are discussed below for both cases).

\begin{figure}[t]
\begin{center}
\includegraphics[width=0.46\textwidth]{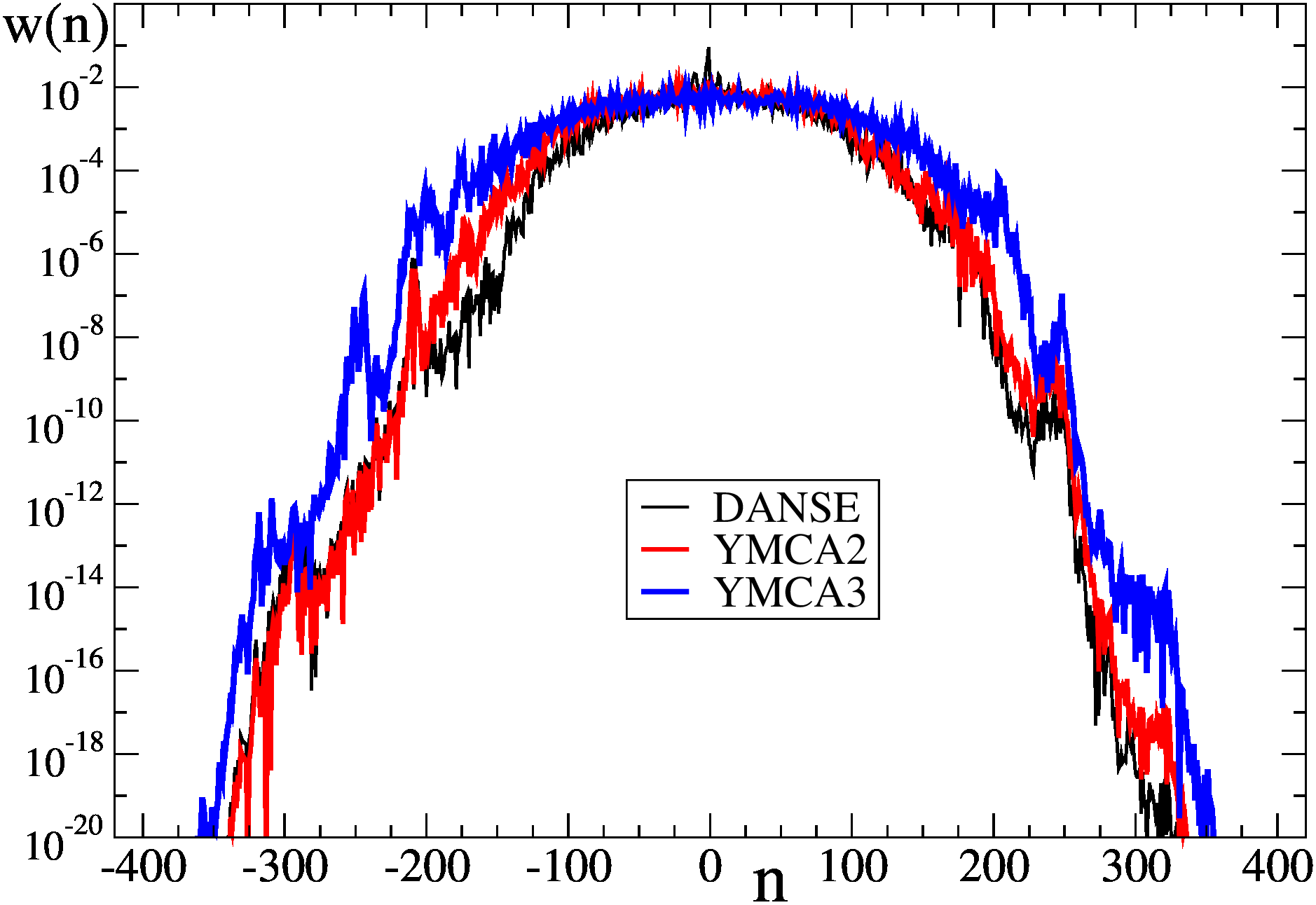}
\end{center}
\vglue -0.3cm
\caption{\label{fig5}
Probability distributions $w(n)$ are shown for  
DANSE (black curve), YMCA2 (red curve) and 
YMCA3 (blue curve) at $\beta=2, W=4$ and $t=10^7$;
probabilities are averaged over 20 disorder realizations.
}
\end{figure}

In Fig.~\ref{fig3} we show directly the probability distribution
over lattice sites $w(n) = \sum_{\mu=1}^{N_C} \mid{\psi^{\mu}_{n}}\mid^2/N_C$ 
at different moments of time for YMCA3 case with $\beta=2$. 
There is a formation of quasi-plateau distribution
which size $\Delta n$ increases with time. Outside of plateau there are
probability tails which drop exponentially
with the site number that corresponds to exponentially
localized Anderson modes of linear problem.

The distributions $w(n)$ at largest reached time $t=10^7$ and 
different values of nonlinearity $\beta$ are shown in Fig.~\ref{fig4}
The width of the above quasi-plateau size $\Delta n$ increases with $\beta$
being approximately $\Delta n  \approx 220, 320, 440$
for $\beta =1, 2, 4$ respectively. These $\Delta n$ values are much larger 
than the Anderson localization length $\ell \approx 6$.
Also the corresponding nonlinear frequency width 
$\Delta \omega \sim 1/\Delta n \ll 1/\ell$ becomes significantly smaller than
a typical frequency spacing between modes inside localization length $\ell$.
Due to these reasons we can argue that the numerical results show an asymptotic spearing
of wavepacket of YM colors. 

\begin{table}
\begin{tabular}{|c|c|c|c|}
\hline
&   \multicolumn{3}{|c|}{$\alpha$ }\\
   \hline
$\beta$&	DANSE	&YMCA2&	YMCA3\\
   \hline
1&	$0.26 \pm 0.02$	&$0.297 \pm 0.020$  &$0.316 \pm 0.010$\\
2&	$0.317 \pm 0.010$	&$0.327	\pm 0.020$ &$0.363 \pm 0.010$\\
4& 	$0.371 \pm 0.020$	&$0.378	\pm 0.020$ &$0.406 \pm 0.010$\\
\hline
\end{tabular} 
\caption{\label{tab1} Exponent $\alpha$ of growth of second moments $\sigma_{1,2} \propto t^{\alpha}$
for DANSE, YMCA2, YMCA3 models at different values of nonlinearity $\beta$ at $W=4$;
the values of exponent are obtained by a fit in the time interval
$2 \leq \log_{10} t \leq 7$ for data averaged over 20 disorder realisations;
initial states have colors located close to each other with $R_C=1$. }
\end{table}

\begin{figure}[t]
\begin{center}
\includegraphics[width=0.46\textwidth]{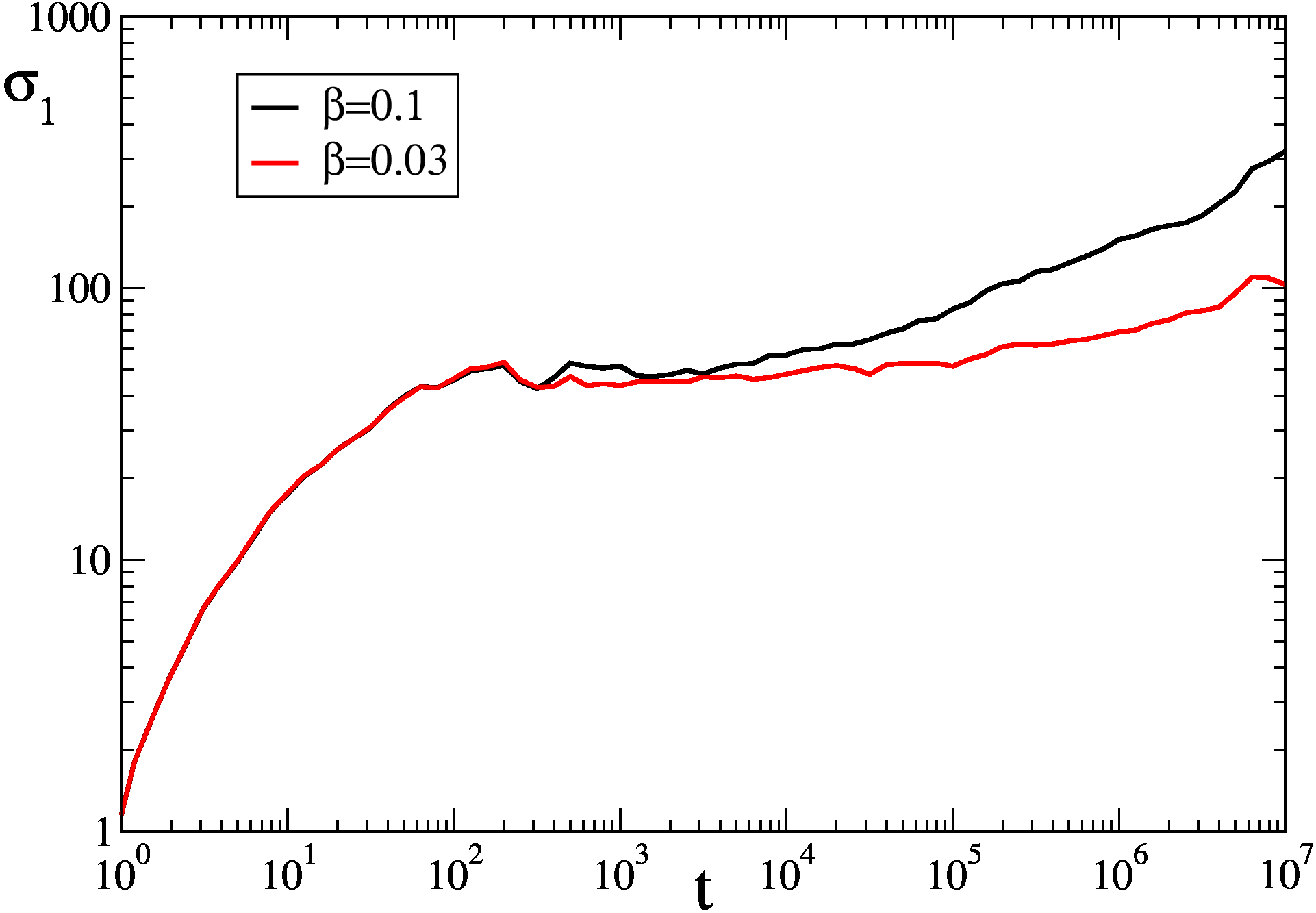}
\end{center}
\vglue -0.3cm
\caption{\label{fig6}
Time evolution of the second moment $\sigma_1$ of probability distribution, defined in the text, 
for YMCA3 model with 3 colors for  
$\beta=0.1$ (black curve), 
$\beta=0.03$ (red curve) at $W=4$. 
At initial time the 3 color packets are located at 3 different sites $n=-1,0,1$.
The results are shown for 10 disorder realizations and 
logarithmic equidistant intervals of time. 
%In a time interval $2 \leq \log_{10} \le 7$ a fit gives: 
%$\alpha = 0.16 \pm 0.03$ for $\beta = 0.1$,
% $\alpha = 0.067 \pm 0.020$ for $\beta = 0.03$.unstable
}
\end{figure}

The comparison of probability distributions
for DANSE, YMCA2, YMCA3 models is shown in Fig.~\ref{fig5}
for fixed $\beta, W$ and $t=10^7$. The  most broad spreading
corresponds to YMCA3 case. This is in a qualitative agreement with an
expectation that, similar to Hamiltonian (\ref{eq_3colorham}),
there is an exact degeneracy of linear color eigenmodes modes 
so that here chaos is present even in the limit
of very small $\beta$ similar to the situation discussed
in \cite{chis2,mulansky2} 
(of course, this assumes that initial 
state have a close location of 3 colors so that
degenerate linear modes are well populated,
see discussion below).

According to the results of Figs.~\ref{fig1},~\ref{fig2} 
the growth of $\sigma_1, \sigma_2$ at large times is well described by an algebraic 
function of time with the exponent $\alpha$. The values of $\alpha$,
obtained from the fit for time range $100 \leq t \leq 10^7$
are given in Table~\ref{tab1} for DANSE, YMCA2, YMCA3 models.
For DANSE at $\beta =1$ the obtained value of $\alpha$ is a bit smaller
than the one reported at \cite{dls2008}
with $\alpha = 0.306 \pm 0.002$. We attribute this difference to a
different number of realisations and longer time range used in \cite{dls2008}.
We also should note that the spreading is rather slow in time
and thus very long time simulations and large number 
of realizations are required to obtain accurate
values of $\alpha$. Formal statistical errors reported here and in \cite{dls2008}
are relatively small but the contribution
of certain systematic effects, related to slow transitions between 
localized linear modes, may give more significant corrections
to formal statistically averaged $\alpha$ values.
From Table~\ref{tab1} we see a moderate increase
of $\alpha$ for higher $\beta$ values.
We also find that YMCA3 and YMCA2 models have
a moderately higher values of $\alpha$ compared to DANSE case.
We attribute this to a stronger chaos for YM colors
compared to DANSE. Indeed, YM colors have additional color
degrees of freedom that are supposed to generate a stronger chaos
thus facilitating deconfinement and spreading of YM fields.
However, due to the above points related to a slow
spreading process further more advanced studies are required
to firmly state if $\alpha$ is independent, or not, of $\beta, W$
and number of colors $N_C$.

\subsection{Confinement and localization of YM colors}

Above we discussed the cases with moderate strength of interactions of colors given by $\beta$.
It is natural to expect that at small $\beta \ll \beta_c \sim 1$ the Anderson localization 
is preserved and fields remain localized in space. Indeed, the numerical results 
reported for DANSE \cite{dls2008} indicate that localization is preserved at small
$\beta=0.1; 0.03$. At the same time we note that in this limit
the effects of slow processes like the Arnold diffusion \cite{chirikov1979,chivech}
are still possible with a very slow spreading of very small fraction of
probability via tiny chaotic layers. The mathematical results
are not able to clarify the behavior in this regime (see e.g. \cite{fishman,wang}).

\begin{figure}[t]
\begin{center}
\includegraphics[width=0.46\textwidth]{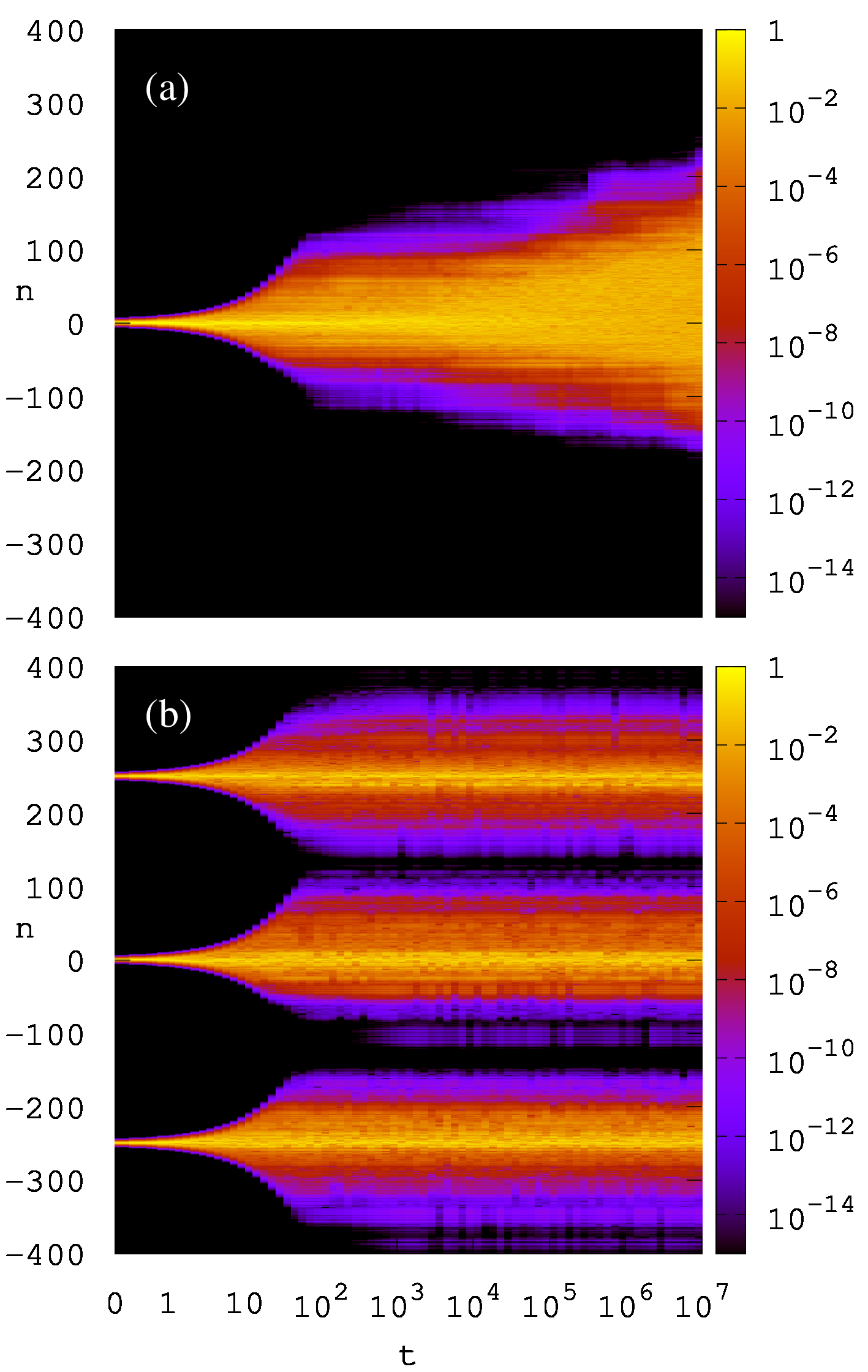}
\end{center}
\vglue -0.3cm
\caption{\label{fig7}
Probability distribution as a function of time $t$ for YMCA3 model at $\beta =2$, $W=4$,
initial positions of 3 colors are $n=-1, 0, 1$ with $R_C=1$ (a);
$n=-250, 0, 250$ with $R_C=250$ (b) for one disorder realisation; 
color bar shows probability $w(n)$ of YM color fields. 
}
\end{figure}

\begin{figure}[t]
\begin{center}
\includegraphics[width=0.46\textwidth]{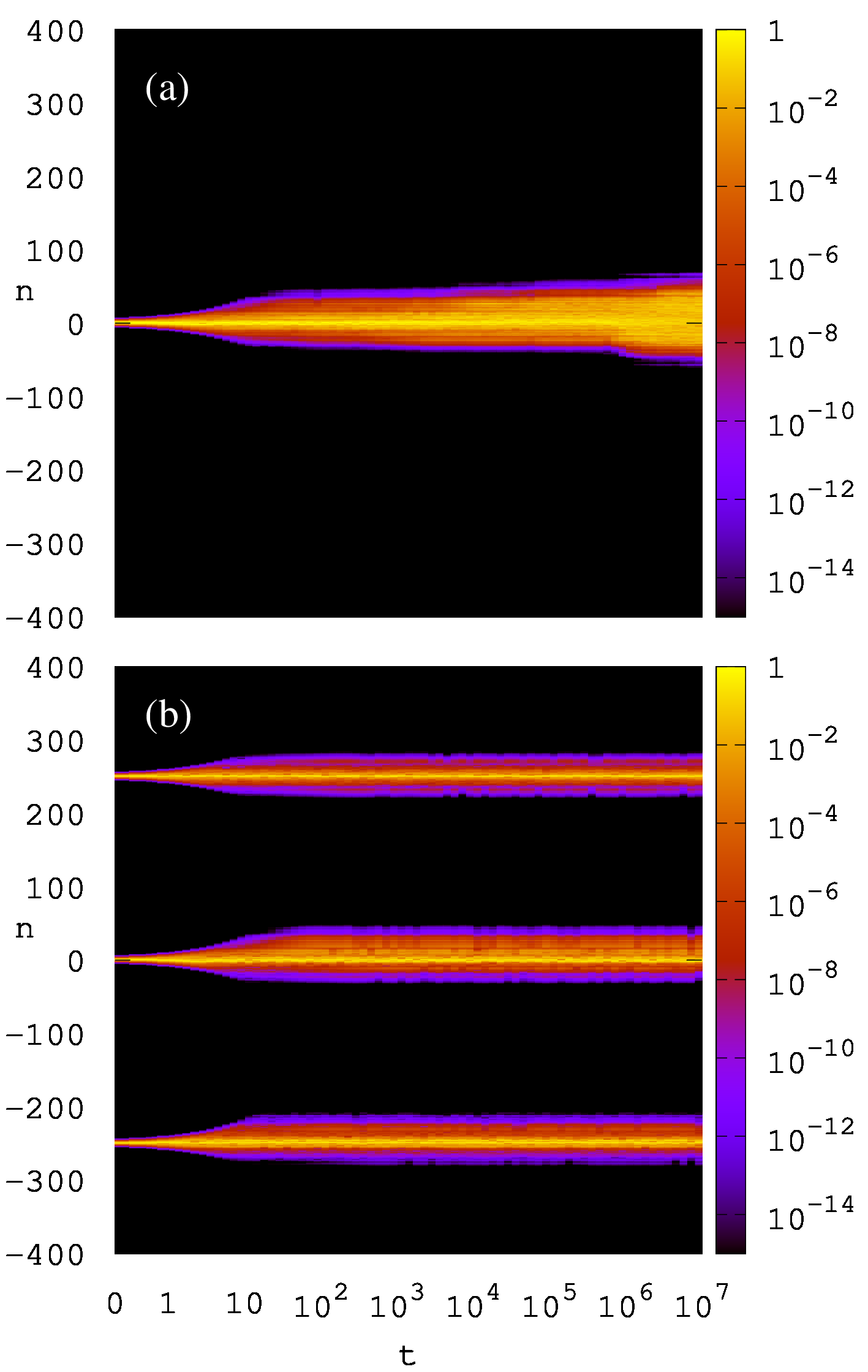}
\end{center}
\vglue -0.3cm
\caption{\label{fig8}
Same as in Fig.~\ref{fig7} but for $W=8$.
}
\end{figure}

\begin{figure}[t]
\begin{center}
\includegraphics[width=0.46\textwidth]{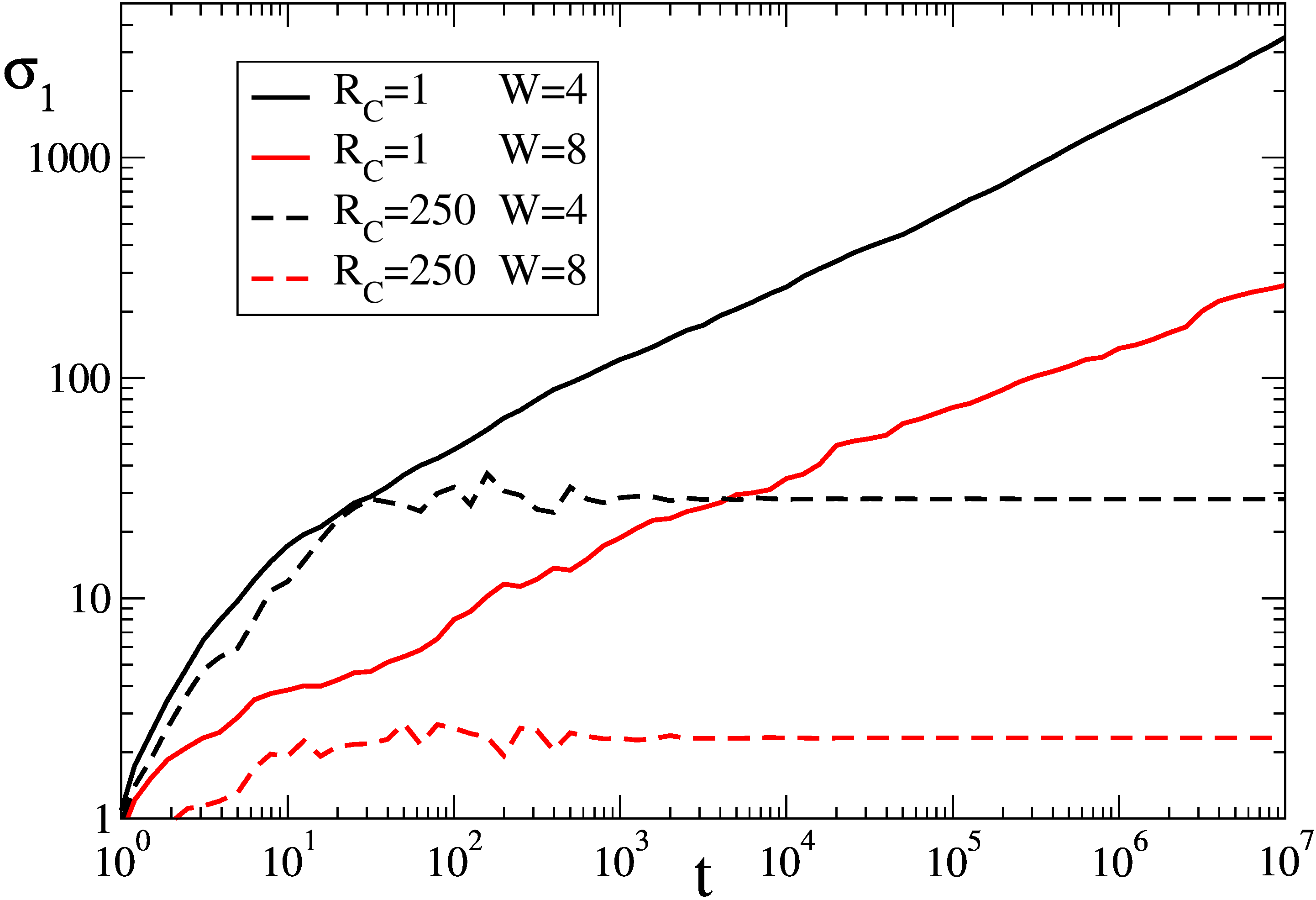}
\end{center}
\vglue -0.3cm
\caption{\label{fig9}
Time dependence of $\sigma_1(t)$ for YMCA3 at $\beta=2$ 
for initial distance between color positions:
$R_C=1$ (full black curve for 20 disorder
realisations at $W=4$ same as in Fig.~\ref{fig1});
$R_C=250$   (dashed black curve for one disorder
realisation at $W=4$ same as in Fig.~\ref{fig7});
$R_C=1$ (full red curve for 10 disorder
realisations at $W=8$);
$R_C=250$   (dashed red curve for one disorder
realisation at $W=8$ same as in Fig.~\ref{fig8}).
}
\end{figure}

For YMCA3 case at such small values of nonlinearity $\beta=0.1; 0.03$ 
we show the time dependence of second moment $\sigma_1$ in Fig.~\ref{fig6}
Here the second moment $\sigma_1$ remains substantially smaller 
compared to $\beta=1, 2, 4$ cases shown in Fig.~\ref{fig1}.
However, a slow increase of $\sigma_1$ at very large times $t >10^5$ is not
excluded. We attribute this to a degeneracy
of linear eigenmodes which, similar to the case of YMCA3 Hamiltonian (\ref{eq_3colorham}),
leads to a high fraction of chaotic phase space even for $\beta \rightarrow 0$,
as discussed in  \cite{chis2,mulansky2} for 3 colors
(we note that for Hamiltonian with 2 colors (\ref{eq_3colorham}) 
there is no chaos in the limit of small
$\beta$  but only a significant energy exchange between two colors 
\cite{chis2}).  Thus, a slow spreading at very large times 
for YMCA3 case may take place  due to
frequency degeneracy present for color fields initially located on a 
distance $R_C < \ell$. The effect of very slow Arnold diffusion
\cite{chirikov1979,chivech} can be also present for a small fraction
of global probability.  

The interesting point is that the above exact frequency degeneracy
is present only if initial color packets are close to each other.
In the opposite case with their initial significant separation
on a distance $R_C \gg \ell$ the effective nonlinear interactions
between colors drop exponentially with $R_C$ due to 
localization of linear eigenmodes. In addition
the frequencies of eigenmodes populated for such 
packets with large separation $R_C \gg \ell$
and statistically different and have no exact degeneracy
in contrast to the case with $R_C < \ell$.
Thus for $R_C \gg \ell$ we argue
that this case corresponds to  
a very small effective interactions with
$\beta_{eff} \propto \beta \exp(-2R_C/\ell) \ll 1$
and that the color
YM fields remain confined and localized.
This is confirmed by the results shown in Figs.~\ref{fig7},~\ref{fig8}
were we compare close and distant location of initial 
color packets. We have clear deconfinement and spreading
for $R_C=1$ (for $W=8$, $\beta=2$, $R_c=1$
the fit gives $\alpha = 0.30 \pm 0.02 $ being smaller than the value
at $W=4$ in Table~\ref{tab1}).
In contrast, for $R_C \gg \ell$ there is  confinement and 
localization of YM color fields. The increase of disorder strength from $W=4$ 
in Fig.~\ref{fig7} to $W=8$ in Fig.~\ref{fig8} gives 
at $R_C=250$ a strong enhancement of
localization of color YM fields.  For distant initial positions
of color fields $R_C =250$ the second moment $\sigma_1(t)$
shows absolutely no growth with time as it is shown in Fig.~\ref{fig9}. 

It is interesting to note that the situation 
with localization-delocalization of color YM fields
reminds those of a quantum problem of two interacting particles
coherently propagating in a disorder potential
and being localized if separated 
by a distance being larger than 
a one-particle localization length
(see e.g. \cite{dlstip,imrytip,frahmtip}).

\subsection{Simple estimates for spreading exponent of YM colors}

Here we present simple estimates for the spreading exponent $\alpha$
of the second moment growth $\sigma_{1,2} \propto t^\alpha$.
Following the approach described in \cite{dls1993,garcia} 
it is useful to rewrite Eq.~(\ref{eq_ym}) in the basis of 
eigenstates of the linear system at $\beta=0$.
The transformation from  lattice representation to
eigenstate basis reads 
$\psi^{\mu}_{n} = \sum_m Q^{\mu}_{nm} C^{\mu}_m$ for each color $\mu$.
Then the time evolution Eq.~(\ref{eq_ym}) takes the form:
\begin{equation}
i {{\partial C^{\mu}_{m}} \over {\partial {t}}}
=\epsilon_{m} C^{\mu}_{m}
+ \beta \sum_{\mu' \neq \mu} \sum_{{m_1}{m_2}{m_3}}
U^{\mu'}_{{m}{m_1}{m_2}{m_3}}
C^{\mu}_{m_1}C^{{\mu'}*}_{m_2}C^{\mu'}_{m_3}
\label{eq_egnbasis}
\end{equation}
where $\epsilon_{m}$
are the eigenenergies of linear system
being the same for all colors.
The transitions between linear eigenmodes
take place only due to the nonlinear $\beta$-term
with the transition matrix elements
$U^{\mu'}_{m m_1 m_2 m_3} = \sum_n (Q^{{\mu}}_{nm})^{-1} Q^{\mu'}_{nm_1} (Q^{\mu'}_{n m_2})^*  Q^{\mu}_{nm_3}$.
Due to the exponential localization of linear eigenstates the sum over each $m$-index
in (\ref{eq_egnbasis})
contains about $\ell$ terms.

In \cite{dls1993} it was argued that in the assumption that
there is a plateau of size $\Delta n$ with random coefficients 
of approximately equal amplitudes and random signs or phases
and zero amplitudes outside the plateau. Then the population of states outside of plateau
should go with the rate $\Gamma \sim |C|^6 \sim 1/(\Delta n)^3$ on nearby sites
on a distance $\ell$. This gives a diffusion rate 
$D \sim \ell^2 \Gamma \sim \ell^2/(\Delta n)^3 \sim (\Delta n)^2/t$
leading to the growth
$(\Delta n)^2 \sim \sigma_{1,2} \sim t^\alpha$ with the spreading exponent
$\alpha=2/5$. 

There are also other type of arguments leading to the same exponent $\alpha=2/5$.
In fact the time evolution of (\ref{eq_egnbasis}) represents 
the nonlinear field dynamics
involving many random frequency components describing a continuous
chaotic flow. The spreading $\Delta n$ in time is very slow
and its Lyapunov exponent $\lambda$ at given $ \Delta n$ is
given by the nonlinear frequency $\lambda \sim \delta \omega \sim \beta /\Delta n$ \cite{dls1993,garcia}.
It is well established that for such a continuous chaotic flows with many frequency components
the diffusion rate $D$ is related with the Lyapunov exponent $\lambda$,
or typical nonlinear frequency $\delta \omega$, by the relation 
established in \cite{chirikov1969,rechester}: $D \sim \lambda^3 \sim (\delta \omega)^3$.
This relation was well confirmed for the Chirikov typical map 
which represents a generic model of such continuous chaotic flows \cite{typmap}
(see also recent work \cite{kurchan}).
Since for (\ref{eq_egnbasis}) we have
$\delta \omega \sim \beta/\Delta n$ this gives us
$D \propto 1/(\Delta n)^3 \propto (\Delta n)^2/t$ and thus
the spreading exponent is $\alpha=2/5$ in agreement with the estimate given at \cite{dls1993}.
We note that for spreading in a disorder potential in higher dimension $d >1$ this approach gives 
the spreading $(\Delta n)^2 =R^2 \sim t^{\alpha}$ with 
the exponent $\alpha = 2/(3d+2)$ with $\alpha=1/4$ for $d=2$ 
(here $R$ is a 1D wavepacket size) \cite{garcia}.

Another estimate of $\alpha$ was proposed in \cite{basko}
on the assumption that the transition rate is given by the Fermi golden rule
as in linear equations of quantum mechanics. This gives
$\Gamma \propto |C|^4 \sim 1/(\Delta n)^2$ and leads to
$\alpha =1/2$,

More complicated estimate arguments were pushed forwards at \cite{flach}
leading to the value $\alpha=1/3$. 

There are various physical arguments behind each of estimates described above.
However, the time evolution of nonlinear YM fields in presence of disorder
is a rather complicated problem. The obtained numerical values of the spreading exponent
are found to be approximately in the range $0.3 \leq \alpha \leq 0.4$.
Further numerical studies are required, with longer times evolution
and larger number of disorder realisations,
to determine more exactly the exponent value.

\subsection{YM color breathers?}

The mathematical proof given in \cite{aubry}
guaranties that nonlinear classical Hamiltonian lattices have generic solutions 
called discrete breathers. 
They represent time-periodic nonlinear field localized, usually exponential, in space.
Such breathers find a variety of applications as discussed in \cite{flachbr1}.
It was shown that breathers exist also for the DANSE model with and without
disorder \cite{flachbr2,politi}.
Usually the breaths appear at a strong nonlinearity of self-interacting field
that effectively creates a solution similar to an impurity
energy level outside of energy band in quantum mechanics.
For the YM color fields (\ref{eq_ym}) nonlinearity appears only due to
interactions of different colors. We suppose that the breather solutions
still can exist for the YM color dynamics on a discrete lattice.
However, the verification of this conjecture requires
further studies which are outside of the scope of this work.

%{\it Discussion.} 
\section{Discussion} 
\label{sec4}

The dynamics of classical homogeneous Yang-Mills color fields
and its chaotic properties have been investigated and well understood 
about 2 decades ago (see e,g \cite{matinyan1,chis1,matinyan2,chis2}).
Here we analyzed the spacial aspects of classical YM color fields
and properties of their propagation in disorder potential in 1D.
In absence of interactions of YM fields the color wavepackets
are confined and exponentially localized by disorder
similar to the Anderson localization of electron transport induced by
disorder \cite{anderson,imry,montambaux,mirlin}. The interactions of YM fields
leads to deconfinement of colors which, above a certain interaction threshold,
spread subdiffusively over the whole disordered lattice.
The exponent of this algebraic spreading is found to be approximately in a range
of $0.3 < \alpha <0.4$ being similar to the value found for
the DANSE model \cite{dls1993,dls2008,flach} and
observed in experiments on cold atoms Bose-Einstein condensate
spreading in a disordered optical lattices \cite{inguscio2}.
Compared to the DANSE model we show that YM color fields
can be deconfined and delocalized only when color component
remain close to each other. In contrast separated color wavepackets
remain confined and localized by disorder.
We expect that the obtained results for classical YM color field dynamics in a disorder
potential will be also useful for the problem of YM fields deconfinement
in the full quantum problem.

\begin{acknowledgments}
This research has been partially supported through the grant
NANOX $N^\circ$ ANR-17-EURE-0009 (project MTDINA) in the frame 
of the {\it Programme des Investissements d'Avenir, France}.
\end{acknowledgments}

%\clearpage
%\pagebreak[4]
%%%%%%%%%%%%%%%%%%%%%%%%%%%%%%%%%%%%%%%%%%%%%%%%%%%%%%%%%


\begin{thebibliography}{99}

\bibitem{ym} C.N.~Yang, and R.L.~Mills,
             {\it Conservation of isotopic spin and isotopic gauge invariance},
             Phys. Rev. {\bf 96}, 191 (1954).
\bibitem{polyakov1} A.M.~Polyakov,
             {\it Particle spectrum in quantum field theory}, JETP Lett. {\bf 20}, 194 (1974) 
             [Pis'ma Zh. Eksp. Teor. Fiz. {\bf 20}, 430 (1974)].
\bibitem{polyakov2} A.M.~Polyakov,
             {\it Isomeric states of quantum fields}, Sov. Phys. JETP {\bf 41(6)}, 988 (1976)
              [Zh. Eksp. Teor. Fiz. {\bf 68}, 1975 (1975)].
\bibitem{polyakov3} A.M.~Polyakov,
             {\it Compact gauge fields and the infrared catastrophe}, 
             Phys. Lett. B {\bf 59}, 82 (1975).
\bibitem{polyakov4} A.A.Belavin, A.M.~Polyakov, A.S.~Schwartz, and Yu.S.~Tyupkin,
             {\it Pseudoparticle solutions of the Yang-Mills equations}, 
              Phys. Lett. B {\bf 59}, 85 (1975).
\bibitem{vainstein} A.I.~Vainstein, V.I.~Zakharov, V.A.~Novikov, and M.A.~Shifman,
              {\it ABC of instantons}, Sov. Phys. Usp. {\bf 25}, 195 (1982).
\bibitem{shuryak2002} D.M.~Ostrovsky, G.W.~Carter, and E.V.~Shuryak,
               {\it Forced tunneling and turning state explosion in pure Yang-Mills theory},
               Phys. Rev. D {\bf 66}, 036004 (2002).
\bibitem{matinyan1} S.G.~Matinyan, G.K.~Savvidi, and N.G.~Ter-Arutunyan-Savvidi,
              {\it Classical Yang-Mills mechanics. Nonlinear color oscillations}, 
               Sov. Phys. JETP {\bf 53(3)}, 421 (1981) 
               [Zh. Eksp. Teor. Fiz.  {\bf 80}, 830 (1981)].
\bibitem{chis1} B.V.~Chirikov, and D.L.Shepelyanskii,
              {\it Stochastic oscillations of classical Yang-Mills fields},
              JETP Lett. {\bf 34}, 163 (1981) 
              [Pis'ma  Zh. Eksp. Teor. Fiz. {\bf 34(4)}, 171 (1981)].
\bibitem{matinyan2} S.G.~Matinyan, G.K.~Savvidi, and N.G.~Ter-Arutyunyan-Savvidi, 
               {\it Stochasticity of classical Yang-Mills mechanics and 
               its elimination by using the Higgs mechanism},
               JETP Lett. {\bf 34}, 590 (1981) 
               [Pis'ma Zh. Eksp. Teor. Fiz. {\bf 34(11)}, 613 (1981)].
\bibitem{chis2} B.V.~Chirikov, and D.L.Shepelyanskii,
              {\it Dynamics of some homogeneous models of classical Yang-Mills fields},
               Sov. J. Nucl. Phys. {\bf 36(6)}, 908 (1982) [Yad. Fiz. {\bf 36}, 1563 (1982)].
\bibitem{chirikov1979} B.V.~Chirikov, 
              {\it A universal instability of many-dimensional oscillator systems},
               Phys. Rep. {\bf 52}, 263 (1979).
\bibitem{lichtenberg} A.~Lichtenberg, and M.~Lieberman, 
              {\it Regular and chaotic dynamics}, Springer, NY (1992).
\bibitem{arnold}  V.~Arnold, and A.~Avez, 
               {\it Ergodic problems in classical mechanics}, 
               Benjamin, NY (1968).
\bibitem{sinai} I.P.~Cornfeld, S.V.Fomin, and Y.G.~Sinai, 
               {\it Ergodic theory}, Springer-Verlag, NY (1982).
\bibitem{ymclass1} D.~Berenstein, and D.~Kawai,
               {\it Smallest matrix black hole model in the classical limit},
                  Phys. Rev. D {\bf 95}, 106004 (2017).
\bibitem{ymclass2} T.~Akutagawa, K.~Hashimoto, T.~Sasaki, and R.~Watanabe,
               {\it Out-of-time-order correlator in coupled harmonic oscillators},
                J. High Energ. Phys. {\bf 2020}, 13 (2020).
\bibitem{ymclass3} G.~Savvidy,
                {\it Maximally chaotic dynamical systems},
                 Annals of Physics {\bf 421}, 168274 (2020).
\bibitem{shuryak1980} E.V.~Shuryak,
               {\it Quantum chromodynamics and the theory of superdense matter},
               Phys. Reports {\bf 61}, 71 (1980).
\bibitem{olesen} P.~Olesen, 
                {\it Confinement and random fluxes},
                Nucl. Phys. B {\bf 200(FS4)}, 381 (1982).
\bibitem{kirzhnits} S.M.~Apenko, D.A.~Kirzhnits, and Yu.E.~Lozovik,
                  {\it Dynamical chaos, Anderson localization, and confinement},
                   JETP Lett. {\bf 36(5)}, 213 (1982) 
                  [Pis'ma Zh. Eksp. Teor. Fiz. {\bf 36(5)}, 172 (1982)].
\bibitem{shuryak1993} E.V.~Shuryak, J.J.M.~Verbaarschot,
                {\it Random matrix theory and spectral sum rules
                 for the Dirac operator in QCD},
                  Nucl. Phys. A {\bf 560}, 306 (1993).
%\bibitem{greensite} J.~Greensite,
%                {\it The confinement problem in lattice gauge theory},
%                Prog. Part. Nucl. Phys. {\bf 51}, 1 (2003).
\bibitem{anderson} P.W.~Anderson,
                 {\it Absence of diffusion in certain random lattices},
                   Phys. Rev. {\bf 109}, 1492 (1958).
\bibitem{imry} Y.~Imry, {\it Introduction to mesoscopic physics},
                   Oxford University Press, Oxford UK (2002).
\bibitem{montambaux} E.~Akkermans, and G.~Montambaux,
               {\it Mesoscopic physics of electrons and photons},
               Cambridge University Press, Cambridge UK (2007).
\bibitem{mirlin} F.~Evers, and A.D.~Mirlin,
                 {\it Anderson transitions},
                 Rev. Mod. Phys. {\bf 80}, 1355 (2008).
\bibitem{dls1993} D.L.~Shepelyansky,
                   {\it Delocalization of quantum chaos by weak nonlinearity},
                 Phys. Rev. Lett. {\bf 70}, 1787 (1993).
\bibitem{molina} M.I.~Molina,
                 {\it Transport of localized and extended excitations 
                  in a nonlinear Anderson model},
                 Phys. Rev. B {\bf 58}, 12547 (1998).
\bibitem{dls2008} A.S.~Pikovsky, and D.L.~Shepelyansky, 
                 {\it Destruction of Anderson localization by a weak nonlinearity}, 
                   Phys. Rev. Lett. {\bf 100}, 094101 (2008).
\bibitem{flach2009} Ch.~Skokos, D.O.~Krimer, S.~Komineas, and S.~Flach,
                  {\it Delocalization of wave packets in disordered nonlinear chains},
                  Phys. Rev. E {\bf 79}, 056211 (2009).
\bibitem{mulansky} M.~Mulansky, and A.~Pikovsky, 
                  {\it Energy spreading in strongly nonlinear disordered lattices}, 
                  New J. Phys. {\bf 15}, 053015 (2013).
\bibitem{flach} T.V.~Lapteva, M.I.~Ivanchenko, and S.~Flach, 
                 {\it Nonlinear lattice waves in heterogeneous media}, 
                 J. Phys. A: Math. Theor. 47: 493001 (2014).
\bibitem{garcia} I.~Garcia-Mata, and D.L.~Shepelyansky, 
                 {\it Delocalization induced by nonlinearity in systems with disorder}, 
                 Phys. Rev. E {\bf 79}, 026205 (2009).
\bibitem{flachkg} Ch.~Skokos, and S.~Flach,
                  {\it Spreading of wave packets in 
                   disordered systems with tunable nonlinearity},
                  Phys. Rev. E {\bf 82}, 016208 (2010).
\bibitem{ermann} L.~Ermann, and D.L.~Shepelyansky, 
                 {\it Destruction of Anderson localization by nonlinearity in 
                  kicked rotator at different effective dimensions}, 
                  J. Phys. A: Math. Theor. {\bf 47}, 335101 (2014). 
\bibitem{flach2019} I.~Vakulchyk, M.V.~Fistul, and S.~Flach, 
                   {\it Wave packet spreading with disordered nonlinear 
                    discrete-time quantum walks}, 
                   Phys. Rev. Lett. {\bf 122}, 040501 (2019).
\bibitem{skokos} B.~Many Manda, B.~Senyange, and Ch.~Skokos,
                  {\it Chaotic wave-packet spreading in two-dimensional 
                   disordered nonlinear lattices},
                  Phys. Rev. E {\bf 101}, 032206 (2020).
\bibitem{segev} T.~Schwartz, G.~Bartal, S.~Fishman, and M.~Segev, 
                 {\it Transport and Anderson localization 
                 in disordered two-dimensional photonic lattices},
                 Nature (London) {\bf 446}, 52  (2007).
\bibitem{lahini} Y.~Lahini, A.~Avidan, F.~Pozzi, M.~Sorel, R.~Morandotti, 
                 D.N.~Christodoulides, and Y.Silberberg,
                 {\it Anderson localization and nonlinearity in one-dimensional 
                 disordered photonic lattices},
                  Phys. Rev. Lett. {\bf  100}, 013906 (2008).
\bibitem{inguscio1} J.E.~Lye, L.~Fallani, M.~Modugno, 
                   D.S.~Wiersma, C.~Fort, and M.~Inguscio,
                   {\it Bose-Einstein condensate in a random potential},
                   Phys. Rev. Lett. {\bf 95}, 070401 (2005).
\bibitem{inguscio2} E.~Lucioni, B.~Deissler, L.~Tanzi, G.~Roati, M.~Zaccanti, M.~Modugno,  
                    M.~Larcher, F.~Dalfovo, M.~Inguscio, and G.~Modugno,
                    {\it Observation of subdiffusion in a disordered interacting system},
                      Phys. Rev. Lett. {\bf 106}, 230403 (2011).
\bibitem{deconf1} P.~Petreczky, {\it Lattice QCD at non-zero temperature},
                  J. Phys. G: Nucl. Part. Phys. {\bf 39}, 093002 (2012).
\bibitem{deconf2} O.~Philipsen, {\it The QCD equation of state from the lattice},
                  Prog. Part. Nucl. Phys. {\bf 70}, 55 (2013).
\bibitem{deconf3} U.~Reinosa, J.~Serreau, M.~Tissier, and N.~Wchebor,
                   {\it Deconfinement transition 
                    in SU(2) Yang-Mills theory: a two-loop study},
                      Phys. Rev. D {\bf 91}, 045035 (2015).
\bibitem{kramer1993} B.~Kramer, and A.~MacKinnon, 
                    {\it Localization: theory and experiment},
                    Rep. Prog. Phys. {\bf 56}, 1469 (1993).
\bibitem{basko} D.~Basko,
                 {\it Kinetic theory of nonlinear diffusion in 
                a weakly disordered nonlinear Schr\"odinger chain 
                 in the regime of homogeneous chaos},
                Phys. Rev. E {\bf 89}, 022921 (2014).
\bibitem{chivech} B.V.~Chirikov, and V.V.Vecheslavov,
                     {\it Arnold diffusion in large systems},
                  JETP Am. Inst. Phys. {\bf 85(3)}, 616 (1997) 
                  [Zh. Eksp. Teor. Fiz. {\bf 112}, 1132 (1997)].
\bibitem{fishman} S.~Fishman, Y.~Krivopalov, and A.~Soffer, 
                  {\it On the problem of dynamical localization in the nonlinear 
                   Schr\"odinger equation with a random potential},
                  J. Stat. Phys. {\bf 131}, 843 (2008).
\bibitem{wang} J.~Bourgain and W.-M.~Wang, 
                {\it Quasi-periodic solutions of
                 nonlinear random Schr\"odinger equations},
                J. Eur. Math. Soc. {\bf 10}, 1 (2008).
\bibitem{ermannnjp} L.~Ermann, and D.L.~Shepelyansky,
                  {\it Quantum Gibbs distribution from dynamical thermalization 
                    in classical nonlinear lattices},
                   New J. Phys. {\bf 15}, 12304 (2013).
\bibitem{mulansky2} M.~Mulansky, K.~Ahnert, A.~Pikovsky, and D.L.~Shepelyansky, 
                   {\it Strong and weak chaos in 
                   weakly nonintegrable many-body Hamiltonian systems}, 
                   J. Stat. Phys. {\bf 145}, 1256 (2011).
\bibitem{dlstip} D.L.~Shepelyansky,
                {\it Coherent propagation of two interacting particles in a random potential},
                  Phys. Rev. Lett. {\bf 73},  2607 (1994).
\bibitem{imrytip} Y.~Imry,
                {\it Coherent propagation of two interacting particles in a random potential},
                   Europhys. Lett. {\bf 30(7)}, 405 (1995).
\bibitem{frahmtip} K.M.~Frahm,
                  {\it Eigenfunction structure and scaling of 
                  two interacting particles in the one-dimensional Anderson model},
                  Eur. Phys. J. B {\bf 89}, 115 (2016).
\bibitem{chirikov1969} B.V.~Chirikov,
                  {\it Research concerning the theory of nonlinear resonance and stochasticity}, 
                  Preprint N 267, Institute of Nuclear Physics, Novosibirsk (1969)
                   [English Transl., CERN Trans. 71 - 40, Geneva, October (1971)].
\bibitem{rechester} A.B.~Rechester, M.N.~Rosenbluth, and R.B.~White,
                   {\it Calculation of the Kolmogorov entropy for 
                    motion along a stochastic magnetic field},
                   Phys. Rev. Lett. {\bf 42}, 1247 (1979).
\bibitem{typmap} K.M.~Frahm and D.L.~Shepelyansky, 
                 {\it Diffusion and localization for the Chirikov typical map}, 
                 Phys. Rev. E {\bf 80}, 016210 (2009).
\bibitem{kurchan} T.~Goldfriend, and J.~Kurchan,
                   {\it Quasi-integrable systems are slow to thermalize 
                    but may be good scramblers},
                    Phys. Rev. E {\bf 102}, 022201 (2020).
\bibitem{aubry} R.S.~MacKay, and S.~Aubry,
                {\it Proof of existence of breathers for time-reversible 
                 or Hamiltonian networks of weakly coupled oscillators},
                  Nonlinearity {\bf 7}, 1623 (1994).
\bibitem{flachbr1} S.~Flach, and A.V.~Gorbach,
                   {\it Discrete breathers - advances in theory and applications},
                   Phys. Reports {\bf 467}, 1 (2008).
\bibitem{flachbr2} G.~Kopidakis, S.~Komineas, S.~Flach, and S.Aubry,
                   {\it Absence of wave packet diffusion in disordered nonlinear systems},
                   Phys. Rev. Lett. {\bf 100}, 084103 (2008).
\bibitem{politi} S.~Iubini, and A.~Politi,
                 {\it Chaos and localization in the discrete nonlinear Schr\"odinger equation},
                 arXiv:2103.11041[nlin.CD] (2021).

\end{thebibliography}
\end{document}